\documentclass[twocolumn,twocolappendix]{aastex631}

\usepackage{romannum}
\usepackage{nicefrac}
\usepackage{xcolor}

\newcommand{\suit}{{\it{SUIT}}}
\newcommand{\degree}{$^{\circ}$}

\graphicspath{{./}{figures/}}
\begin{document}

\title{X-class flare on Dec 31, 2023, observed by the Solar Ultraviolet Imaging Telescope on board Aditya-L1}

\author[0000-0003-2215-7810]{Soumya Roy}
\affil{Inter-University Centre for Astronomy and Astrophysics, Post Bag 4, Ganeshkhind, Pune 411007}\email{soumyaroy@iucaa.in}
\author[0000-0003-1689-6254]{Durgesh Tripathi}
\affil{Inter-University Centre for Astronomy and Astrophysics, Post Bag 4, Ganeshkhind, Pune 411007}
\author[0000-0002-9253-6093]{Vishal Upendran}
\affil{Bay Area Environmental Research Institute, Moffett Field, USA}
\affil{Lockheed Martin Solar and Astrophysics Laboratory, Palo Alto, CA, USA}
\author[0000-0002-7276-4670]{Sreejith Padinhatteeri}
\affil{Manipal Centre for Natural Sciences, Manipal Academy of Higher Education, Karnataka, Manipal- 576104, India}
\author[0000-0001-5707-4965]{A. N. Ramaprakash}
\affil{Inter-University Centre for Astronomy and Astrophysics, Post Bag 4, Ganeshkhind, Pune 411007}
\author[0000-0001-6866-6608]{Nived V. N.}
\affil{Inter-University Centre for Astronomy and Astrophysics, Post Bag 4, Ganeshkhind, Pune 411007}
\author[0000-0003-1406-4200]{K. Sankarasubramanian}
\affil{UR Rao Satellite Centre, Indian Space Research Organisation, Old Airport Road, Vimanapura PO, Bengaluru, 560017, India}
\affil{Center of Excellence in Space Sciences India, Indian Institute of Science Education and Research Kolkata, Kolkata, 741246, India}
\author[0000-0002-3418-8449]{Sami K. Solanki}
\affil{Max-Planck-Institut für Sonnensystemforschung, Justus-von-Liebig-Weg 3, 37077 Göttingen, Germany}
\author[0000-0002-8560-318X]{Janmejoy Sarkar}
\affil{Inter-University Centre for Astronomy and Astrophysics, Post Bag 4, Ganeshkhind, Pune 411007}
\affil{Tezpur University, Napaam, Tezpur, Assam 784028}
\author[0000-0002-1282-3480]{Rahul Gopalakrishnan}
\affil{Inter-University Centre for Astronomy and Astrophysics, Post Bag 4, Ganeshkhind, Pune 411007}

\author[0009-0000-2781-9276]{Rushikesh Deogaonkar}
\affil{Inter-University Centre for Astronomy and Astrophysics, Post Bag 4, Ganeshkhind, Pune 411007}
\author[0000-0001-5205-2302]{Dibyendu Nandy}
\affil{Center of Excellence in Space Sciences India, Indian Institute of Science Education and Research Kolkata, Mohanpur 741246, West Bengal, India}
\affil{Department of Physical Sciences, Indian Institute of Science Education and Research Kolkata, Mohanpur 741246, West Bengal, India}
\author[0000-0003-4653-6823]{Dipankar Banerjee}
\affil{Indian Institute of Space Science and Technology
Valiamala, Thiruvananthapuram - 695 547, Kerala, India}
\affil{Center of Excellence in Space Sciences India, Indian Institute of Science Education and Research Kolkata, Mohanpur 741246, West Bengal, India}

\correspondingauthor{Soumya Roy}

\begin{abstract}

We present the multi-wavelength study of the ejection of a plasma blob from the limb flare SOL2023-12-31T21:36:00 from  NOAA 13536 observed by the Solar Ultraviolet Imaging Telescope (SUIT) on board Aditya-L1. We use SUIT observations along with those from Atmospheric Imaging Assembly (AIA) on board SDO and Spectrometer/Telescope for Imaging X-rays (STIX) on board Solar Orbiter to infer the kinematics and thermal nature of the ejected blob and its connection to the associated flare. The observations show that the flare was comprised of two eruptions. The blob was ejected during the first eruption and later accelerated to velocities over 1500~km~s$^{-1}$ measured at a maximum projected height of $\sim$ 178~Mm from the Sun's surface. The acceleration of the ejected plasma blob is co-temporal with the bursty appearance of the hard X-ray light curve recorded by STIX. Radio spectrogram observations from STEREO-A/WAVES and RSTN reveal type~\Romannum{3} bursts at the same time, indicative of magnetic reconnection. DEM analysis using AIA observations suggests the plasma blob is comprised of cooler and denser plasma in comparison to the ambient corona. To the best of our knowledge, this is the first observation of such a plasma blob in the NUV providing crucial measurements for eruption thermodynamics. 
\end{abstract}
\section{Introduction} \label{sec:intro}
Solar flares are the explosive release of magnetic energy, which are manifested differently across various layers of the solar atmosphere. The variations in their manifestation in different layers may be attributed to the differing physical processes at play. Based on multi-wavelength observations and numerical simulations, a standard model, also known as the``CSHKP" model, for solar flares has been proposed \citep{carmichael64,sturrock66,hirayama74,kopp76}. The corona usually exhibits flare loops filled with evaporated plasma from the chromosphere \citep{neupert67,fletcher11,tripathi04} and supra arcade fans heated up by various mechanisms \citep{svestka,longcope11, reeves17,xie23}, often observed in Extreme Ultraviolet (EUV) and Soft X-ray (SXR) radiation. 

Eruptive flares are often associated with occurrences of filament/prominence eruption and coronal mass ejections (CMEs) \citep{gopalswamy03}. When a CME is associated with a filament eruption, it may exhibit a three-part structure with a bright leading edge, a dark cavity and a bright core consisting of cold material \citep{cremades04}. The prominence eruption also exhibits a characteristic velocity signature, which is known as two-part acceleration. The initial phase, when the prominence material slowly rises in that atmosphere, is called the ``slow-rise phase", followed by the ``fast-rise phase", where the prominence material undergoes strong acceleration \citep{moore_01, chifor06,chifor07,isobe06}. The filament eruptions have also been reported to exhibit pulsating and prolonged hard X-rays during the ``fast-rise phase" due to the reconnection happening in multiple stages \citep{sterling04}. In addition to these, multiple type~\Romannum{3} burst has been linked to fragmented dynamic reconnection \citep{karlicky_2023}.

In this Letter, we report the first limb flare and the associated ejection of a plasma blob observed by the Solar Ultraviolet Imaging Telescope \citep[SUIT;][]{article,ghosh16,suit_main,suit_test_calib} onboard Aditya-L1 \citep{adityal1,tripathi23} in the \ion{Mg}{2}~h filter (NB04). The event was an X5.0 flare with an associated CME, which occurred on 2023 December 31. We use multi-wavelength observations to probe the dynamics and thermal structure of the ejected plasma blob associated with the eruption. 

The remainder of the paper is structured as follows. In \S\ref{sec:obs}, we describe the observations and data used. In \S\ref{sec:res}, we present the data analysis and results. Finally, we summarize and discuss our results in \S\ref{sec:dis}.
\section{Observations and Data} \label{sec:obs}
\begin{figure*}[ht!]
    \centering
    \includegraphics[trim={0.5cm 3cm 0.5cm 3.5cm}, clip, width=1.0\textwidth]{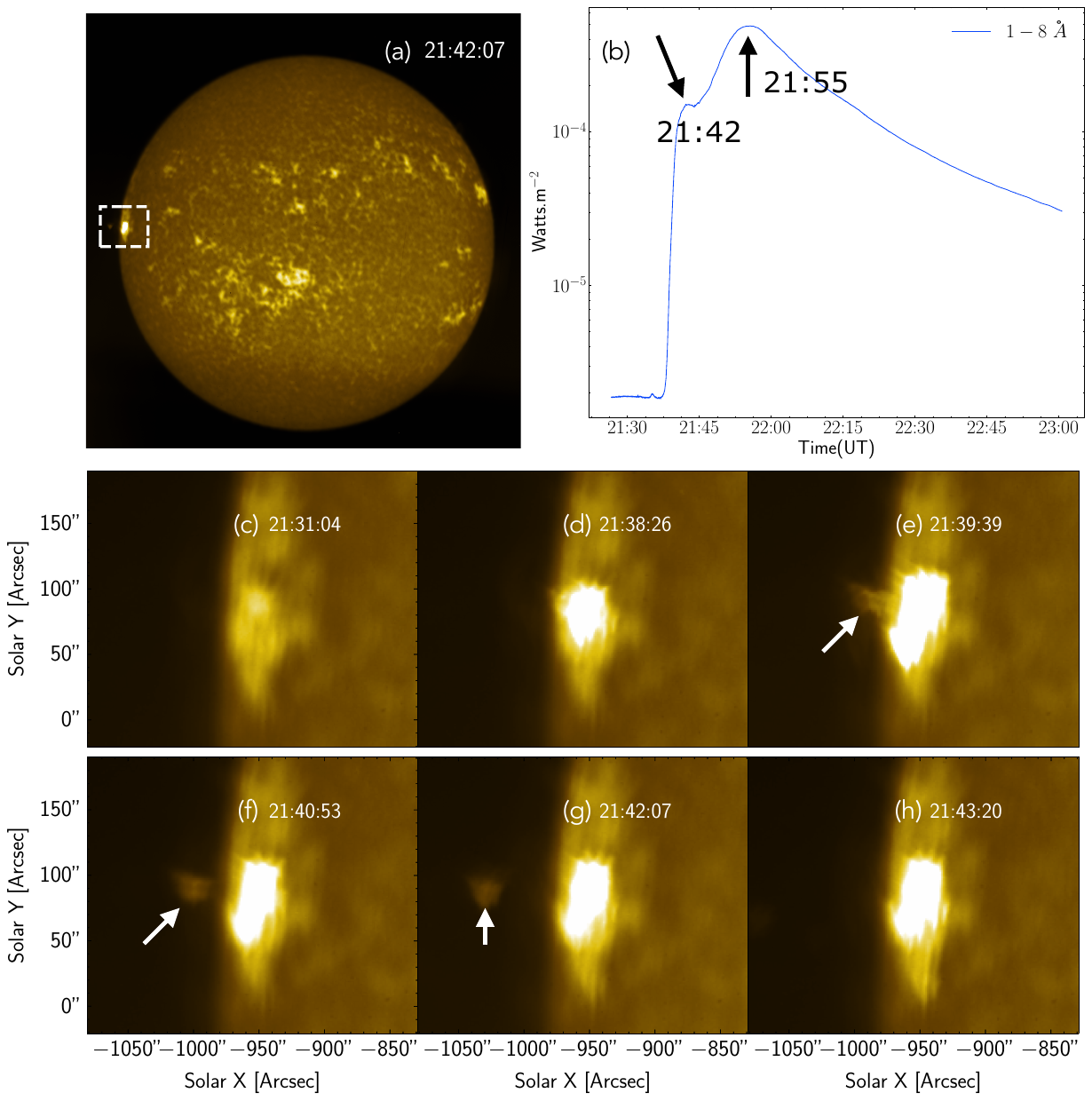}
    \caption{(a) Full disk binned image recorded by {\suit} in the \ion{Mg}{2}~h filter. The over-plotted white dashed box locates the flaring region. (b) {\it GOES} SXR~1{--}8~{\AA} observation of the flare. The flare consists of two soft X-ray peaks, which are marked by two black arrows and their corresponding timing. (c) {--} (h) Time evolution of flare in \ion{Mg}{2}~h and associated ejected plasma blob located with white arrows. The field of view corresponds to the boxed region shown in panel a.}
    \label{fig:dec_flare_suit}
\end{figure*}

The active region NOAA~13536 appeared on the visible side of the disk $\sim$ [-950\arcsec, 60\arcsec] on 31 December 2023 as observed by SUIT (Fig.~\ref{fig:dec_flare_suit}~a) and produced an X5 flare at 21:36 UT that peaked at 21:55~UT in {\it GOES} observations (Fig.~\ref{fig:dec_flare_suit}~b). During normal operations, {\suit} observes the photosphere and the chromosphere of the Sun in the 200{--}400~nm wavelength range and provides full disk and partial disk images in 11 different channels with a pixel size of 0.7{\arcsec} \citep{suit_main}. In addition, it also takes 2k$\times$2k binned images in the NB04 filter centered around the \ion{Mg}{2} h 280.3 nm line with 1~min cadence, which is used for onboard flare detection and localization. At the time of this observation, {\suit} was still in the cruise phase to the Lagrange L1 point, and the onboard flare detection algorithm \citep{flare_det} was not enabled. Therefore, the payload did not get into flare mode. Hence, it only observed the flare in the binned images with a cadence of one minute. 
\begin{figure*}[ht!]
    \centering
    \includegraphics[trim={0cm 3cm 0cm 3.6cm}, clip, width=0.9\linewidth]{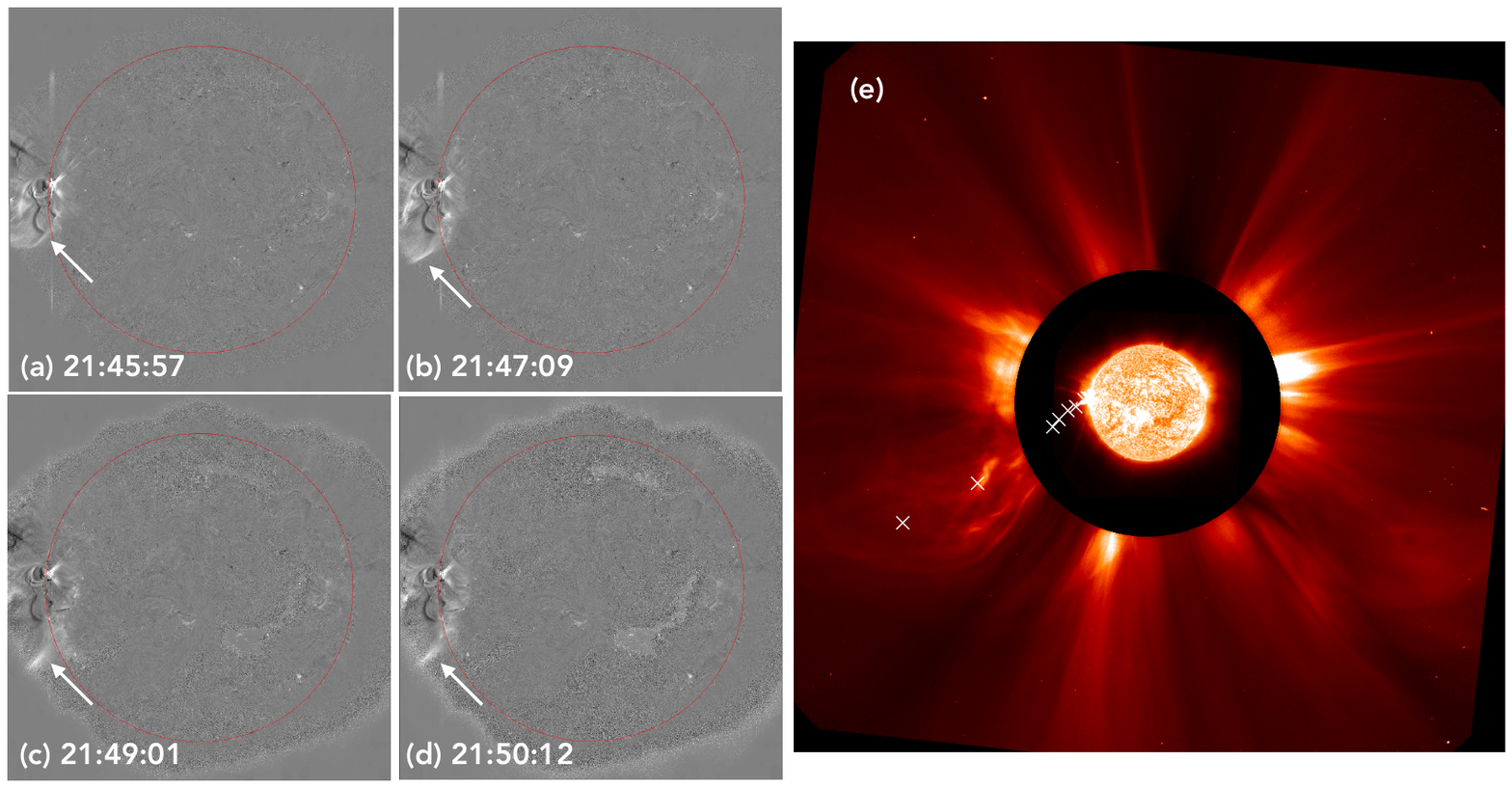}
    \caption{Panels (a) {--} (d) display a sequence of base difference images with respect to 21:30:57 UT in AIA 211~{\AA} showing the propagation of the EUV wave that originated from the flare. White arrows mark the EUV wave. Panel (e) shows the SUVI 304~{\AA} and LASCO/C2 observation of the CME originating from the flare.} 
    \label{fig:cme_obs}
\end{figure*}

The flare was also observed by the Atmospheric Imaging Assembly \citep[AIA;][]{aia} on board the Solar Dynamics Observatory \citep[SDO;][]{sdo}, and the Spectrometer/Telescope for Imaging X-rays \citep[STIX;][]{stix,stix1} on board Solar Orbiter \citep[][]{so}. AIA observes the Sun in seven EUV wavelengths {\it viz.} 171, 94, 131, 171, 193, 211, 304, 335~{\AA}, sensitive to coronal plasma at different temperatures \citep{o'dwyer10,O'Dwyer12}, with a pixel size of 0.6{\arcsec} and cadence of 12~s. It also observes in two continuum channels {\it viz} 1600~{\AA} and 1700~{\AA} with the same pixel size but at a lower temporal resolution of 24~s. In this paper, we use the 171~{\AA} and 1600~{\AA } observations for the kinematic study of the eruption and compare the observations made with SUIT and the six EUV channels (excluding 304~{\AA}) to derive the differential emission measure and thereby temperature maps of the erupting region. 

The flare was also observed with GONG H$\alpha$\citep{gong}. We use the GONG observations to exhibit the similarity of the observed plasma blob in \ion{Mg}{2}~h and H$\alpha$. STIX provides spatially resolved X-ray spectra in the energy range of 4{--}150~keV. We have used full-Sun light curves and STIX images in 4{--}18~keV and 25{--}50~keV for our analysis. In addition we have used radio spectra obtained from STEREO-A/WAVES \citep{swaves} and RSTN-Palehua solar observatory.

The flare was also associated with a CME that was detected by Large Angle Spectrometric Coronagraph Experiment C2 \& C3 \citep[LASCO;][]{lasco}. We have used the C2 observations to study the association of the ejected blob with the CME. There was a propagating EUV wave on the surface of the Sun in the wake of the eruption. The EUV wave was also observed via EUV observations from AIA, the Extreme Ultraviolet Imager \citep[EUVI;][]{euvi} on board Stereo-A \citep{stereo} and Solar Ultraviolet Imager \citep[SUVI;][]{suvi}.

To use SUIT data in conjunction with AIA and STIX, we required a fine co-alignment of these observations. As Aditya-L1 was still in the cruise phase when this observation was recorded, a proper location of the spacecraft was required to make the correction for different vantage points at which the three spacecraft, i.e., Aditya-L1, SDO, and Solar Orbiter, were located. For this purpose, we use the `{\it get\_horizons\_coord}' function available in Sunpy~\citep{sunpy20} to make a query to `JPL HORIZONS' and get the accurate location of Aditya-L1 and Solar Orbiter. We use the obtained location with the `{\it get\_observer\_meta}' function available in Sunpy to generate an updated header for the given spacecraft position. We then co-register and co-align {\suit} NB04 (\ion{Mg}{2}~h) observations with those taken by AIA and STIX. STIX hard X-ray contours are aligned to AIA 1600~{\AA} images and re-projected to AIA perspective. The hard X-ray contours needed to be co-aligned separately as the distance between Solar Orbiter and Sun was greater than 0.75 AU and the aspection solutions available from the SSWIDL pipeline were not reliable. We use the same shifts to the thermal contours also before projecting them to the AIA perspective. We check the alignment of the thermal contours with the emission observed in AIA 193~{\AA} to ensure the reliability of the alignment.

\section{Data Analysis and Results} \label{sec:res}
\subsection{Multi-wavelength evolution the ejected plasma blob and associated flare and CME}\label{sec:multi_wav}
Fig.~\ref{fig:dec_flare_suit}~a displays a full-disk (2k $\times$ 2k) observation recorded by {\suit} in the NB04(\ion{Mg}{2}~h) channel during the impulsive phase of the flare. The flaring region is located with a white box. Panels~c{--}h of Fig.~\ref{fig:dec_flare_suit} displays the sequence of six images corresponding to the box, showing the evolution of the flare. After about 2{--}3 minutes of the start of the flare, we observe an ejection of the plasma blob (located by white arrows) from the flaring site that moves outwards. The GOES light curve shown in Fig.~\ref{fig:dec_flare_suit}~b suggests that there are two flares peaking at $\sim$21:42~UT and $\sim$21:55~UT, suggestive of two eruptions.

Fig.~\ref{fig:cme_obs}.a{--}d displays a sequence of running difference maps obtained from AIA 211~{\AA} observations showing the EUV wave associated with the eruption. White arrows mark the propagating wavefront in all of the panels. 

The associated CME appeared in the LASCO~C2 FoV around $\sim$ 22:04 UT. The speed of the CME as per the SOHO LASCO CME catalogue \citep{lasco_cme_cat} obtained using a linear polynomial fit to the height time plot is 2852~km~s$^{-1}$. Fig.~\ref{fig:cme_obs}e shows SUVI 304~{\AA} observation co-aligned with LASCO~C2. The propagation of the ejected material is marked with white crosses across the SUVI and LASCO FoV.
\begin{figure*}[ht!]
    \centering
    \includegraphics[trim={0cm 0.5cm 0.5cm 0.cm}, clip, width=0.9\textwidth]{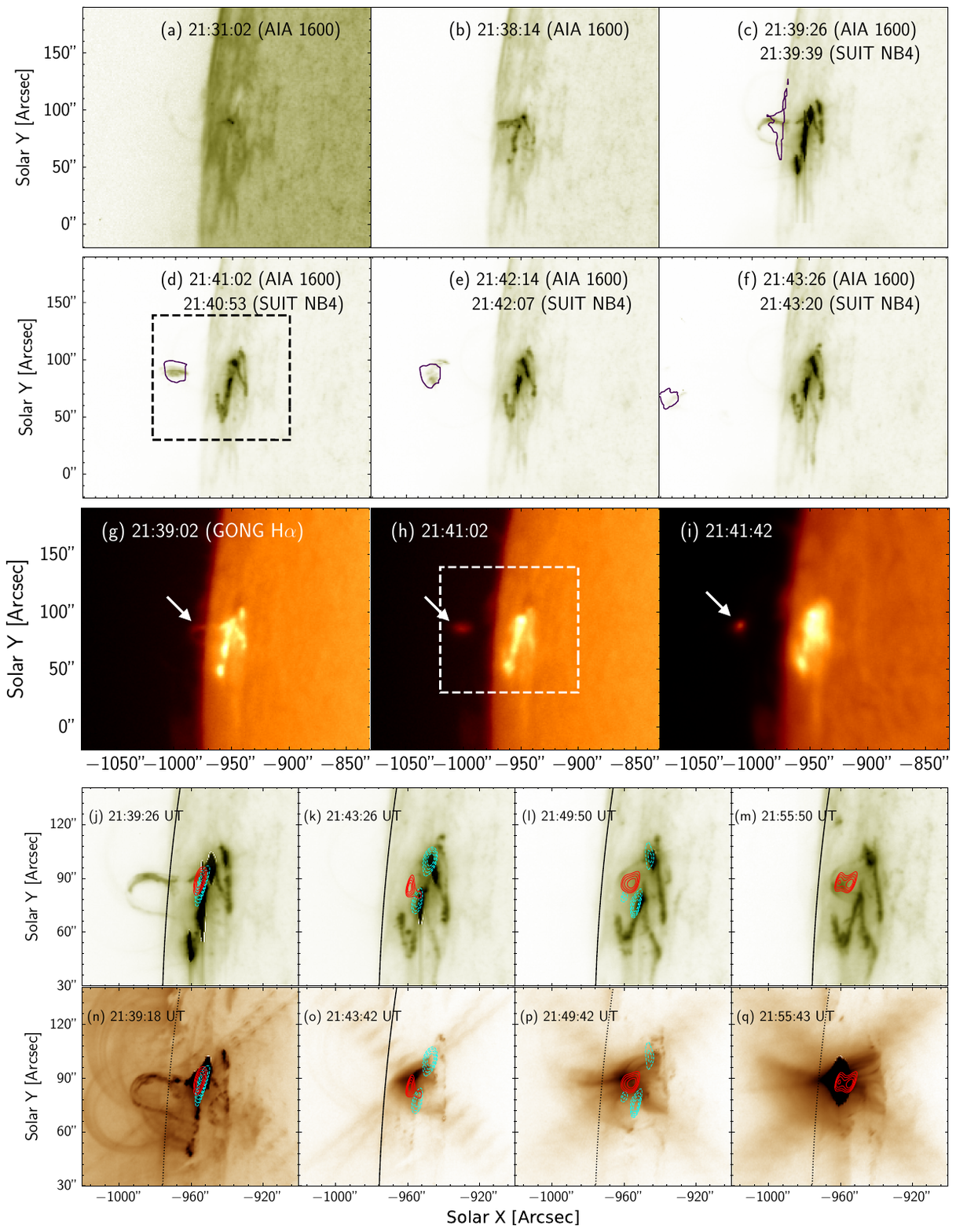}
    \caption{(a){--}(f) Sequence of AIA 1600~{\AA} negative intensity observations. Over-plotted purple contours in (c){--}(f) show the position of the ejected blob in the SUIT NB04 (\ion{Mg}{2} h) observation. (g){--}(i) GONG H$\alpha$ observation sequence of the flaring region as in panel (a). The rising loop observed in AIA 1600~{\AA} (c) is also observed at the same time in H$\alpha$ and marked with a white arrow in panel (g). In the subsequent panels, the ejected plasma blob, as observed in H$\alpha$, is marked with a white arrow. (j){--}(m) Sequence of AIA 1600~{\AA} negative intensity observations from the region marked with the dashed black box in (d) over-plotted with STIX hard X-ray contours (cyan dashed) and soft X-ray contours (red solid). (n){--}(q) Sequence of AIA 193~{\AA} negative intensity observations over-plotted with STIX hard X-ray contours (cyan dashed) and soft X-ray contours (red solid).}
    \label{fig:dec_flare_obs}
\end{figure*}

Fig.~\ref{fig:dec_flare_obs}.a{--}f displays a sequence of negative intensity 1600~{\AA} observations of the eruption. The over-plotted purple intensity contours in panel c{--}f mark the position of the ejected plasma blob from co-aligned and co-registered NB04 (\ion{Mg}{2} h) observations. Correspondingly, we also observe a rising loop structure in AIA 1600~{\AA} (see panel c), which is co-spatial with the ejecting plasma blob as observed in {\suit} NB04 at 21:39 UT, which continues to be observed in subsequent frames. 

The eruption was also captured in H$\alpha$ observations recorded by GONG (see  Fig.~\ref{fig:dec_flare_obs}.g{--i}). The structure of the ejected material observed in H$\alpha$ is strikingly similar to that recorded by the SUIT. We note that while the ejected material has a blob-like structure in both SUIT and H$\alpha$ observations, and they are co-spatial with the structure observed in AIA 1600, the structure seen in the latter is part of an ejected loop. These observations, therefore, may allude to the possibility that while the ejected material is accelerated, probes at different wavelengths are capturing various parts of the same structure depending on their sensitivity. Examining GONG, SUIT and LASCO observations, we believe there are strong suggestions that the observed plasma blob is part of filament material embedded in the CME.

In Fig.~\ref{fig:dec_flare_obs}~j{--m}, we display AIA 1600~{\AA} images over-plotted with STIX 4{--}18~keV (solid red contours) and 25{--}50~keV (dashed cyan contours) at various stages during the evolution of the flare. The FoV of these images corresponds to the boxed regions shown in panels d and h. The two ribbons are clearly visible in panels k{--}m. Fig.~\ref{fig:dec_flare_obs}~n{--q} shows the same hard and soft X-ray contours over-plotted on a sequence of negative intensity AIA 193~{\AA} observation.

Fig.~\ref{fig:dec_flare_obs}~j and n taken during the first eruption, shows the rising loop while the two flare ribbons are not well separated. At this time, the cyan and red contours are co-spatial, implying that the soft and hard X-ray sources are co-spatial. Interestingly, these sources lie at the northern foot-point of the rising loop, along which the cold material is seen to be ejected in {\suit} NB04 observations (see panel c). At later times, during the evolution of the flare, the two ribbons and the flare arcades are well observed as shown in panels k{--}m and o{--}q respectively. The two sets of cyan contours are co-aligned with the two flare ribbons as observed in 1600~{\AA}, whereas the red contours are in between the ribbons and aligned with the EUV loops as observed in 193~{\AA}. This is suggestive of two hard X-ray foot point sources with soft X-ray loops in between. Panels m and q show the AIA 1600~{\AA} and 193~{\AA} observations, respectively, co-aligned with the STIX soft X-ray contours during the peak soft X-ray flux of the second eruption. The hard X-ray flux by this time is too weak to be imaged properly.

\subsection{Kinematic evolution of the ejected plasma blob and its relation with Hard X-rays}\label{sec:kin}
\begin{figure*}[ht!]
    \centering
    \includegraphics[trim = {1cm 0cm 1cm 0.5cm}, clip, width=1.0\textwidth]{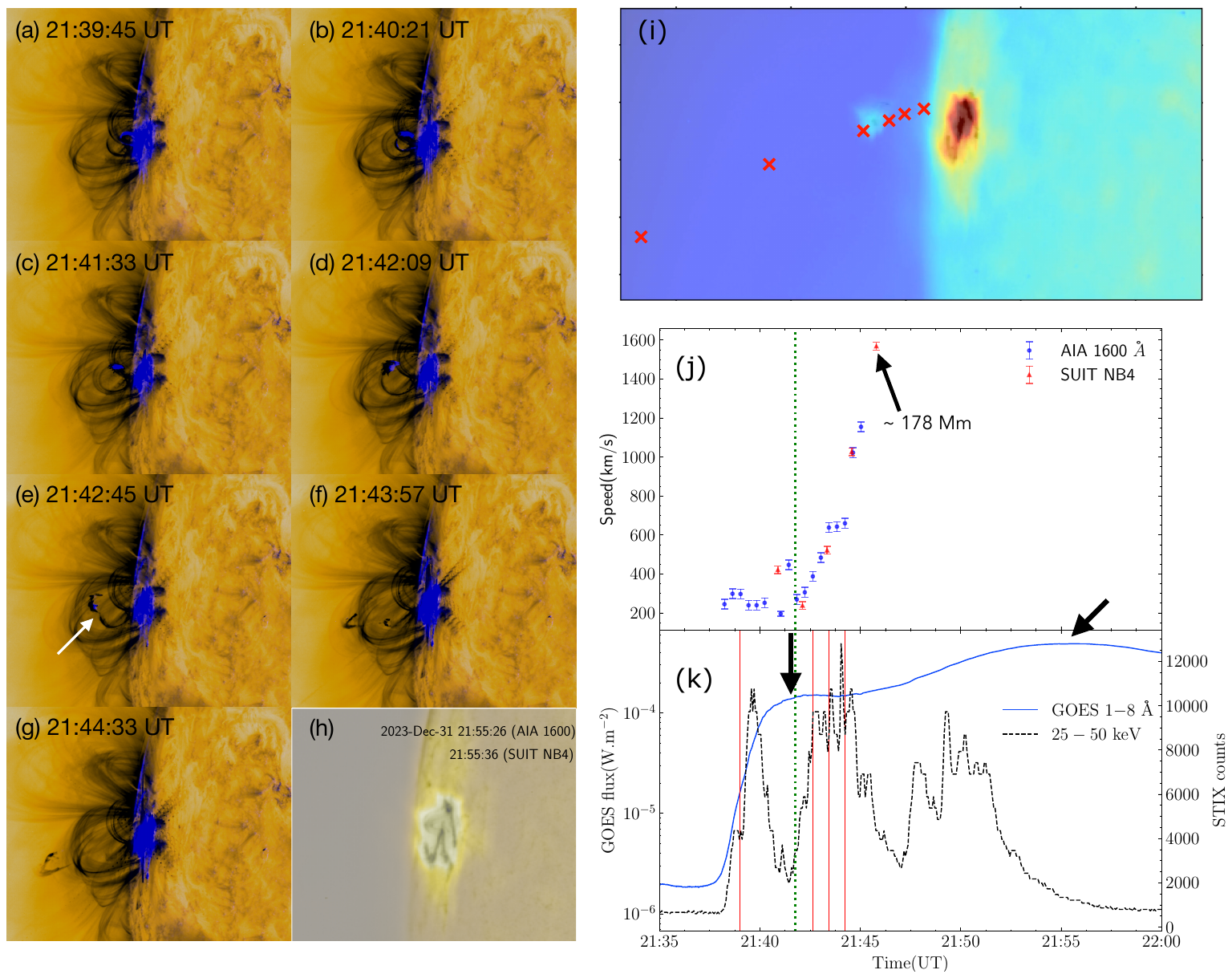}
    \caption{Panels (a){--}(g): the evolution of the event as observed in AIA 1600~{\AA} (blue) and 171~{\AA} (negative intensity). The white arrow in panel e locates the kink. Panel (h):  SUIT NB04 image (yellow) recorded at the flare peak (2nd flare) over-plotted with AIA 1600~{\AA} negative intensity image. Panel (i): SUIT NB04 image recorded at 21:42~UT, over-plotted with red crosses, which mark the leading edge of the ejected plasma blob and are used to study the trajectory of the plasma blob. Panel (j): velocity of the ejected plasma blob as a function of time as obtained using AIA 1600~{\AA} (blue solid circle) and \suit~NB04 (red triangle) observations. The projected height of the blob from the limb at the edge of the SUIT FoV is marked in the panel with the black arrow. Panel (k): GOES 1{--}8~{\AA} flux (blue solid line and left Y-axis) and STIX HXR (25{--}50~keV, black dotted line and right Y-axis). The four solid vertical lines mark the times of panels (a), (e), (f) \& (g). The vertical dotted lines shown in panels (j) \& (k) mark the start of the acceleration of the plasma blob. The two black arrows mark the two soft X-ray peak.}
     \label{fig:dec_flare_vel}
\end{figure*}

Figure~\ref{fig:dec_flare_vel}~a{--}g display a sequence of observations taken in AIA~171~{\AA} (negative intensity) over-plotted with 1600~{\AA} in blue. The sequence of images demonstrates that there are two eruptions. The first eruption is seen at 21:39:45~UT with the foot point brightening in 1600~{\AA} (panel a). From panels b to d, we observe a rising loop in 1600~{\AA} images. The northern arm of the 1600~{\AA} loop corresponds to the plasma blob that is seen in SUIT NB04 images and appears brighter than the rest of the loop. We note that, just after a minute, while the blob keeps its identity, the rest of the loop is no longer seen in 1600~{\AA} (panels c \& d). In panel e, we observe a kink (located by a white arrow) developed in overlying loops observed in 171~{\AA}. After about a minute of the appearance of the kink, a part of the loop is ejected out, as can be seen in panels f and g. An NB04 image combined with 1600~{\AA} native intensity image is shown in panel h taken at the second GOES peak.

We show an NB04 image taken at 21:42~UT in panel i, over-plotted with red crosses corresponding to the leading edge of the ejected blob. We compute the velocity of the plasma blob following this trajectory using NB04 (red triangle) and 1600~{\AA} (blue circle) observations and plot them in panel j. The velocity profiles obtained from SUIT and 1600 are consistent with each other. The velocity is calculated by tracing the leading edge of the ejected material and taking the difference between two successive frames. The NB04 velocity measurements in panel j (red triangles) are calculated by measuring the projected distance between successive red crosses marked in panel i in pixel coordinates. The velocity profile suggests that the plasma blob initially moved with a constant speed of about 300~km~s$^{-1}$ before getting accelerated to velocities more than 1500~km~s$^{-1}$. The error bars are estimated by taking the physical distance of $\mathrm{\sqrt{2}}$ pixels in the respective observations. We note that this is only a lower limit of the uncertainty. The other uncertainties may arise from the errors in determining the leading edge itself, due to the irregular shape of the plasma blob.

\begin{figure*}
    \centering
    \includegraphics[trim={2.5cm .5cm 2.5cm .3cm}, clip, width=0.95\linewidth]{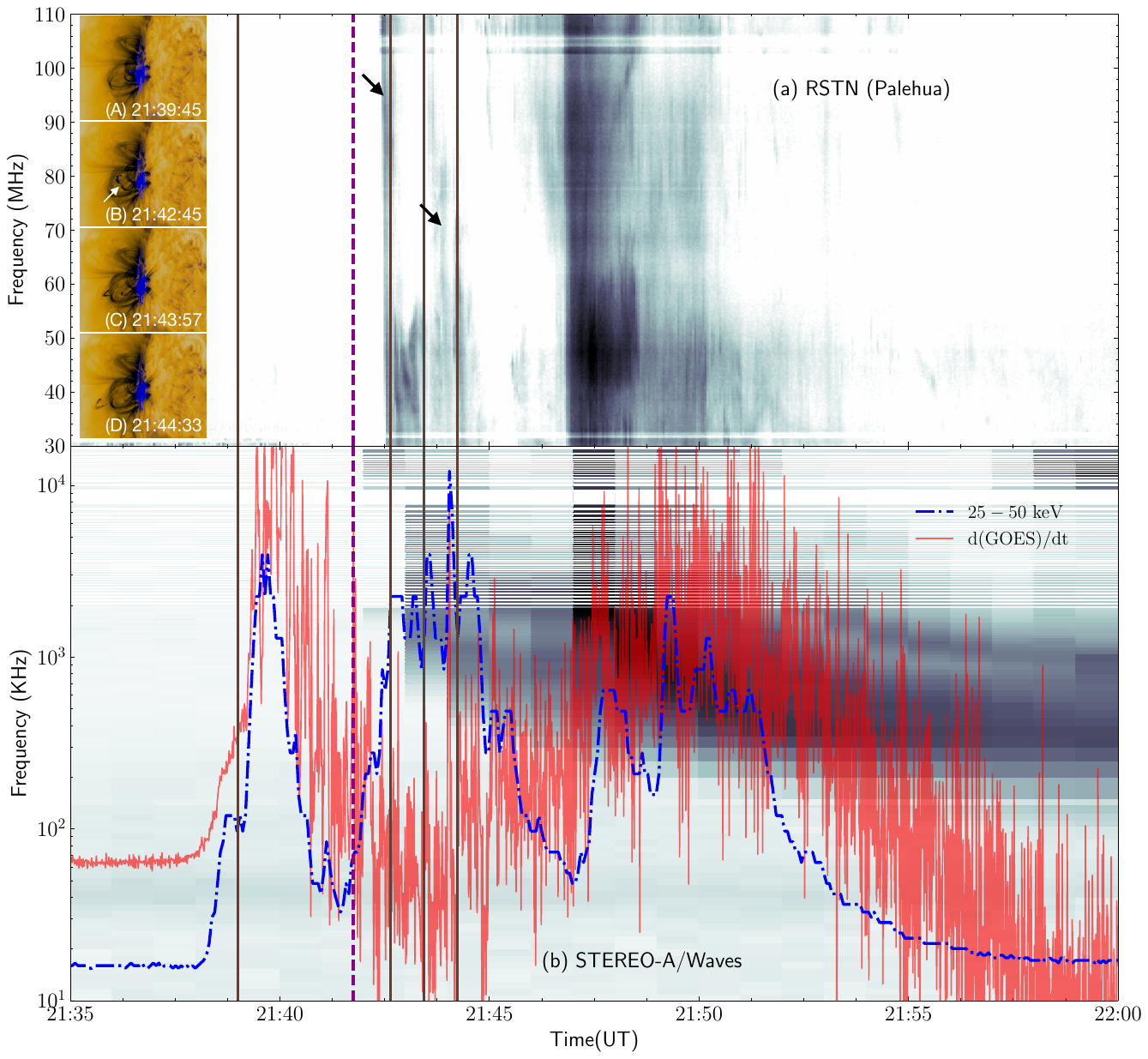}
    \caption{STIX hard X-ray 25{--}50~keV(blue dot-dashed line), and time derivative of GOES 1{--}8~{\AA} (transparent red solid line) over-plotted on RSTN radio spectrogram (panel a) and STEREO-A/WAVES radio spectrogram (panel b). The inset panels A, B, C and D show combined AIA 171 and 1600~{\AA} observations during this time. The brown solid vertical lines are the times of inset panels A, B, C, and D. The magenta dashed vertical line marks the start of the acceleration of the plasma blob, as observed from SUIT.} 
    \label{fig:goes_stix_swaves}
\end{figure*}

Fig.~\ref{fig:dec_flare_vel}~k plots STIX 25{--}50~keV light curve (dashed black line) during the flare showing its bursty nature. There are three distinct hard X-ray activities as evident from the three distinct peaks in the light curve. The over-plotted solid blue curve is the {\it GOES} 1{--}8 {\AA} light curve. The two black arrows mark the two distinct soft X-ray peak observed from GOES. To highlight the sequence of events observed in the AIA images, we plot vertical red lines indicating the times of Fig.~\ref{fig:dec_flare_vel}a, e, f and g. From the plots, it is apparent that the soft X-ray flux reaches a plateau around 21:42~UT and rises again, exhibiting two eruptions from the same active region. The STIX Hard X-ray (HXR) also shows signs of a continued bursty nature, suggestive of dynamic reconnection \citep{kliem_2000} and consistent with hard X-ray observations during the ``fast rise phase" of prominence eruption \citep{sterling04}, as observed in the second and third hard X-ray peak of Fig.~\ref{fig:dec_flare_vel}~k. The acceleration phase of the blob is temporally associated with the second hard X-ray peak which shows a bursty nature. During the same time period successive steady brightening was observed in AIA 1600 and 1700~{\AA}. In addition, in the same time, STIX imaging reveals that majority of the hard X-ray originate at the two foot points (see Fig.~\ref{fig:dec_flare_obs} panel k{--}m). The third hard X-ray peak, while showing similar nature as the second one, cannot directly be associated with the acceleration of the plasma blob as, {by that time it has gone} out of the SUIT FoV and cannot be traced anymore.

To explore the dynamics further, in Fig.~\ref{fig:goes_stix_swaves}, we plot the STIX 25{--}50~keV hard X-ray (blue dot-dashed line) and time derivative of the GOES 1{--}8~{\AA} soft X-ray (solid red line) on top of the radio spectrum obtained from RSTN and STEREO-A/WAVES in $\sim$~30 to 110~MHz and $\sim$~10 kHz to 16.025~MHz frequency respectively. The brown vertical lines mark the times of AIA 1600 and AIA 171 composite observations shown in inset panel A, B, C, and D. The magenta dashed line is the start of the acceleration of the SUIT blob as marked in fig.~\ref{fig:dec_flare_obs}~panel j and k. The `Neupert effect' \citep{neupert68} suggests that $\mathrm{\frac{d}{dt}(f_{SXR})\propto f_{HXR}}$, as the non-thermal energy injected from the precipitating electrons and ions are eventually converted into the thermal energy of the radiating plasma. In fig.~\ref{fig:goes_stix_swaves}, the time derivative of the soft X-ray flux exhibits two distinct peaks and corresponds nicely with the first and third hard X-ray peak observed from STIX hard X-ray (blue dot-dashed line), while the second hard X-ray peak has no counterpart in the time derivative of the soft X-ray. This suggests that the second hard X-ray peak did not have a significant contribution to the thermal plasma, or there was a delayed response of the thermal plasma. However, we note that during the second and third hard X-ray peaks, there are several radio bursts. During the second hard X-ray peak, we locate at least two type~\Romannum{3} radio bursts in the RSTN radio spectra (marked with black arrows in fig.~\ref{fig:goes_stix_swaves}~panel a), which coincides with the acceleration of the SUIT blob. The first radio burst (second brown line) coincides exactly with the time of inset panel Fig.~\ref{fig:goes_stix_swaves}~B, where the development of kink in the loops was noted.

We also note pulsations in the HXR light curve which have been associated with electron injection events along the closed flare loops \citep{collier_24} from STIX observations and dynamic magnetic reconnection \citep{kliem_2000} from 2D MHD simulations. Several type~\Romannum{3} bursts at the same time fit into the picture of dynamic magnetic reconnection injecting energetic electrons into the flares loops (creating the hard X-ray observed in chromosphere) and along the open field lines (creating the type~\Romannum{3} bursts).

\subsection{Thermal structure}\label{sec:therm}

\begin{figure*}[ht!]
    \centering
    \includegraphics[trim = {0cm 0cm 0cm 0cm}, clip, width=1.0\textwidth]{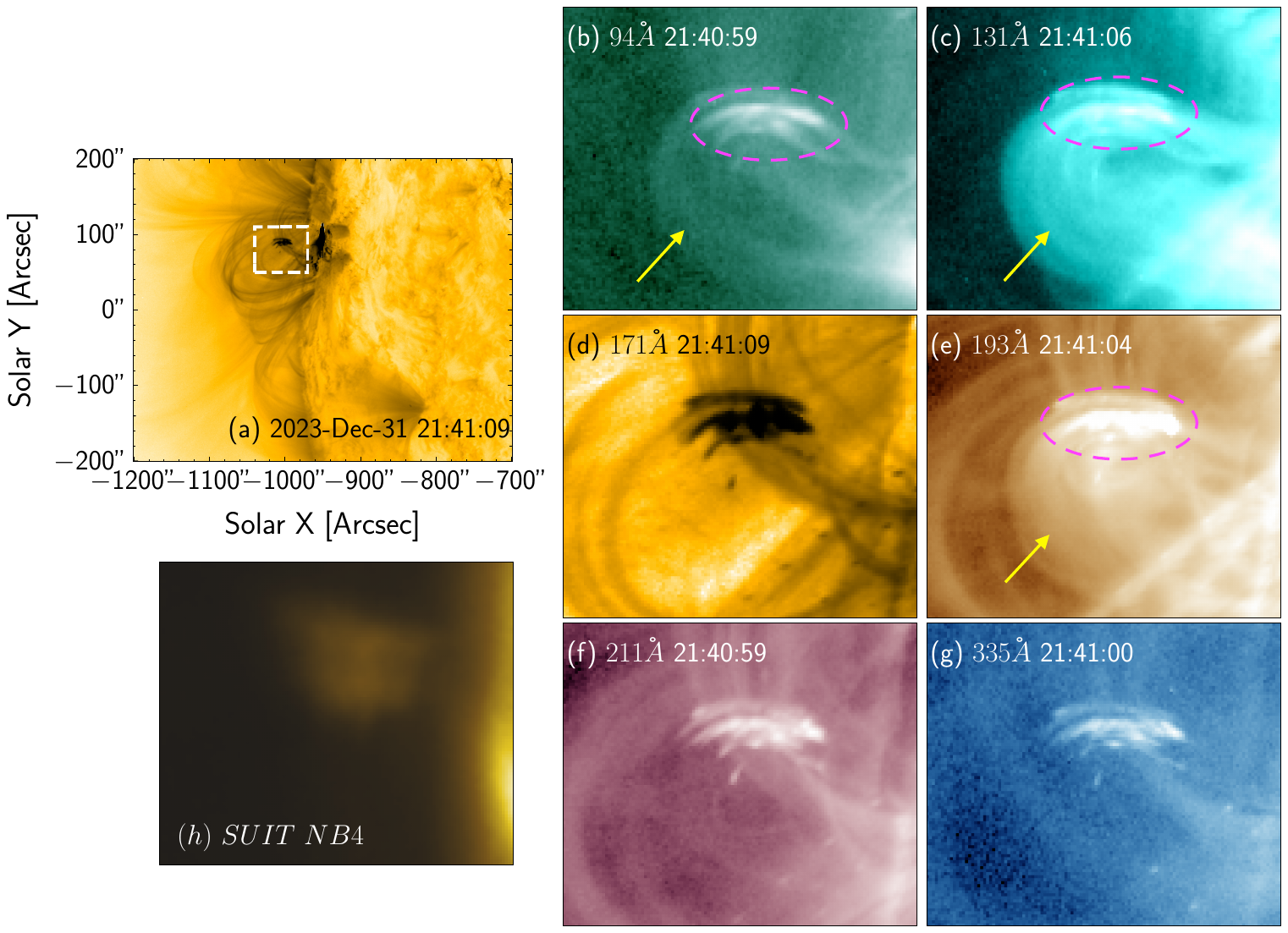}
    \caption{Panel (a) AIA 171~{\AA} observation during the eruption in negative intensity. The white box locates the erupting plasma blob. Panels (b){--}(g): AIA observation in 94, 131, 171, 193, 211, and 335~{\AA} corresponding to the boxed region shown in panel a. Panel (h): {\suit} NB04 observation corresponding to the boxed region in panel a. In panel b,c and e we mark the region of the loop cospatial with the SUIT blob, with magenta dashed ellipse. The same region also appears in the colder channels of AIA in panels d,f and g. The rest of the loop structure appear bright in the hot channels of AIA e.g. 94, 131, and 193. This part of the loops are marked with yellow arrow. This portion of the loop do not appear in the colder channel as seen in panels d, f, and g.}
    \label{fig:dec_flare_aia}
\end{figure*}

\begin{figure}[ht!]
    \centering
    \includegraphics[trim={0.5cm 2.5cm 0.5cm 3.2cm}, clip, width=0.9\linewidth]{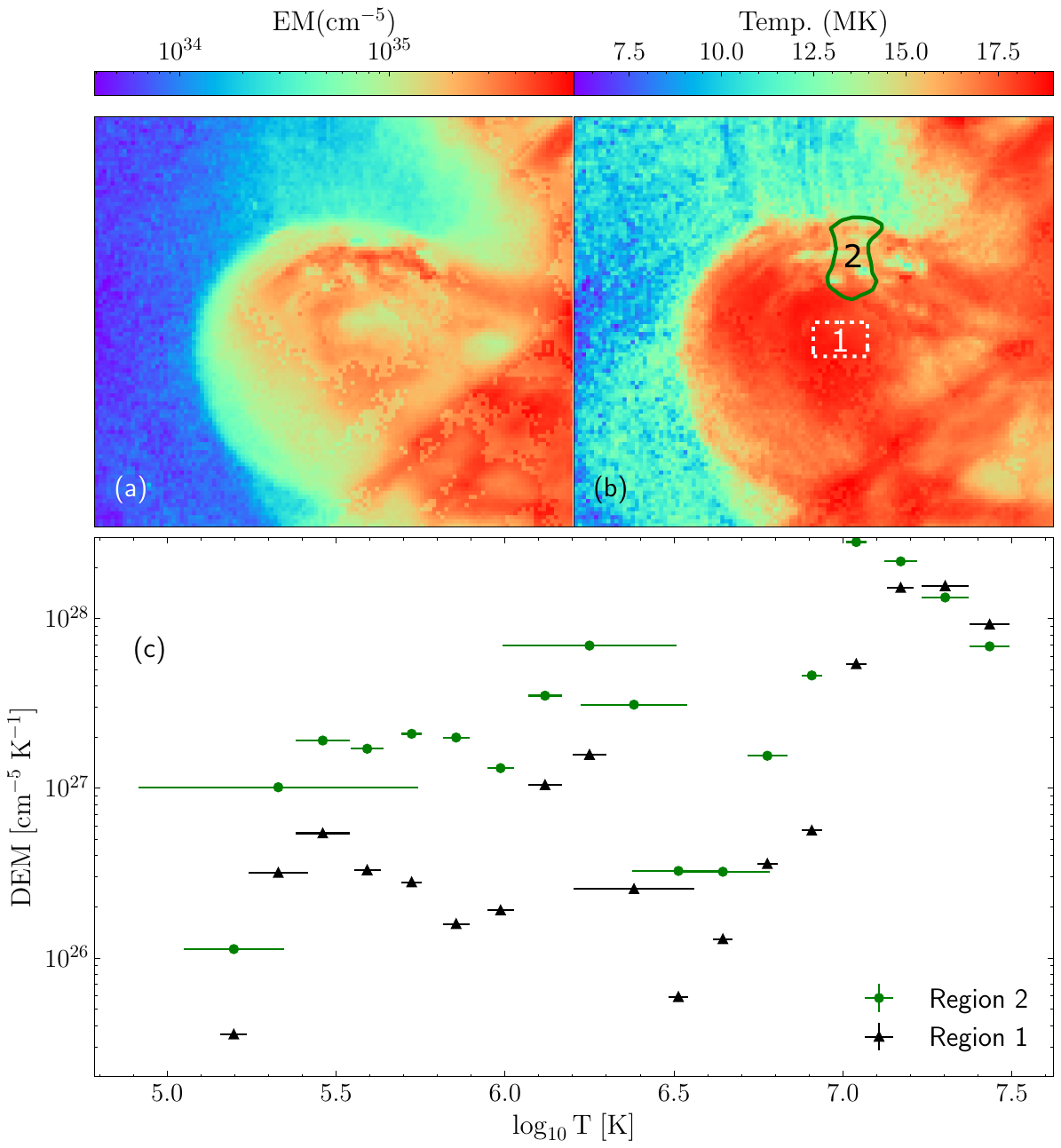}
    \caption{Emission Measure (panel a) and temperature (panel b) maps of the region shown in the boxed region in Fig.~\ref{fig:dec_flare_aia}.a. Panel c plots the DEM as a function of temperature for regions 1 and 2 located in panel b.}
    \label{fig:dec_flare_dem}
\end{figure}

In Fig.~\ref{fig:dec_flare_aia}~a, we show the eruption in AIA 171~{\AA} in negative intensity. The white dashed box locates the region that is shown in panels b{--}g recorded in different AIA channels and panel h as seen in SUIT NB04. Note that panel d is in negative intensity. As can be seen, the ejecting plasma blob is observed in all the AIA channels as well as NB04, demonstrating its multi-thermal nature. The position of the plasma blob region is marked with magenta dashed line in the hot channels of AIA, e.g. 94, 131, and 193~{\AA} in Fig.~\ref{fig:dec_flare_aia}~b,c, and e respectively. We further note that some parts of the loop structure which are visible in hotter channels, e.g., in 94, 131, 193, (marked with yellow arrows in panels b,c, and e) are not seen in 171, 211, 335. Note that at the higher resolution of AIA, the plasma blob looks like parts of field lines and are bright in all six AIA channels. It seems to correspond to the rest of the field lines of the complete loop structure in the hotter channel. In cold channels, e.g. 171, 211, and 335~{\AA} the part of the loop which is visible in the hot channels (marked with yellow arrow in panels b,c, and e) is not visible. This can be explained if the parts of the loop cospatial with the blob is colder and has significantly higher density along the line of sight. The rest of the loop which is not visible in the colder channel is significantly hotter.

To understand the thermal structure of the ejecta, we perform differential emission measure (DEM) analysis using the six coronal channels of AIA. For this purpose, we employ the regularized inversion scheme of \cite{hannah&kontar12} on the field of view located by the white box in Fig.~\ref{fig:dec_flare_aia}~a. The obtained emission measure (EM) and DEM-weighted temperature maps are shown in Fig.~\ref{fig:dec_flare_dem}a~\&~b, respectively, suggesting that the region co-spatial with the plasma blob as seen in NB04 is cooler and denser than the rest of the ejected structure. 

To obtain a further quantitative understanding, we choose two regions shown by the white dotted box (labelled 1) and a solid green contour (labelled 2) in Fig.~\ref{fig:dec_flare_aia}~b. Note that the solid green contour corresponds to the 85\% intensity of the plasma blob observed in NB04. Fig.~\ref{fig:dec_flare_dem}.c plots the DEM obtained on the averaged intensities over these two regions as labelled. The DEM plots demonstrate that while both regions have a similar amount of plasma at higher temperatures, there is a significantly large amount of cooler plasma in region 2. This is a plausible reason for the blob to be seen better in cooler channels of AIA and SUIT NB04. 

\section{Summary and discussion} \label{sec:dis}
In this paper, we discuss the first limb X-class flare (SOL2023-12-31T21:36:00) and the associated ejected plasma blob observed by SUIT, on December 31, 2023. We combined SUIT observations with data from AIA and STIX and performed a comprehensive multi-wavelength study of the evolution of the flare and the associated ejected plasma blob. The eruption was associated with a CME and the plasma blob observed by SUIT is likely filament material embedded in the CME core.

The blob seen in NB04 images closely corresponds to the 1600~{\AA} observations. The velocity profiles obtained using NB04 images and 1600~{\AA} images are identical (Figure.~\ref{fig:dec_flare_vel}.j). The blob reaches to more than $\sim1500$~km~s$^{-1}$ at its maximum height $\sim$~178 Mm from the surface of the Sun, before disappearing from the field of view of SUIT. Thanks to the larger FOV of SUIT, this is for the first time the kinematics of such a blob could be studied to this height in the NUV. 

Our analysis reveals that the flare was associated with two eruptions and with an ejected plasma blob observed prominently in the NB04 filter of SUIT. The blob was also observed in all channels of AIA, demonstrating its multi-thermal nature, which is also revealed by DEM analysis on AIA observations (Fig.~\ref{fig:dec_flare_dem}). To understand the blob emission better, we consider a control sample from the central part of the loop. This comparative analysis tells us that the blob and the loop have similar amount of material at high temperatures, while the blob has larger amount of material at low temperatures, in agreement with the enhanced emission in SUIT NB04 and AIA 1600.

Our observations demonstrate the ejection of cold material into the corona and accelerated outwards through two eruptions. The associated HXR observation from STIX suggests the signatures of dynamic reconnection, the epoch of those matched with the acceleration of the blob. To the best of our knowledge, this is the first such observation in NUV, specifically in any of the \ion{Mg}{2} lines. 

The STIX hard X-ray observation exhibits three distinct hard X-ray peaks during the event, with pulsations in the second and third peaks. The time derivative of the GOES soft X-ray correlates temporally with the first and third hard X-ray peaks. This may indicate that the second hard X-ray peak did not contribute to the thermal plasma efficiently or the thermal plasma exhibits a delayed response. Note that the pulsations observed in the STIX hard X-ray have been associated with multiple variable electron injection into the flare loops \citep{collier_24}, consistent with type~\Romannum{3} bursts observed in the radio spectrogram, as that signifies accelerating electrons along open field lines. The next important step towards understanding the physics of such eruptions is to combine multi-wavelength observations including radio imaging with theoretical modeling. We aim to take it up in the future. 

The observations of the bursty hard X-ray and type~\Romannum{3} radio bursts coinciding with the acceleration phase of the blob and its kinematics measurements at much higher heights demonstrate the uniqueness of SUIT instrument in the understanding of solar eruptions. The SUIT observations highlighted here were carried out during the cruise phase, hence observations were only made in the \ion{Mg}{2}~h channel. Regular full-disk observations from SUIT open up the opportunity to observe similar eruptions in various near-ultraviolet wavelengths, covering mostly photospheric and chromospheric continuum and lines. Combining with various multiwavelength observations, including those from radio, allows us to connect these observations across various layers of the solar atmosphere.

\newpage
\begin{acknowledgements}
We thank the referee for constructive comments. SUIT is built by a consortium led by the IUCAA, Pune, and supported by ISRO as part of the Aditya-L1 mission. The consortium consists of CESSI-IISER Kolkata (MoE), IIA, MAHE, MPS, USO/PRL, and Tezpur University. Aditya-L1 is an observatory class mission, which is funded and operated by the ISRO. The mission was conceived and realised with the help from all ISRO Centres and payloads were realised by the payload PI Institutes in close collaboration with ISRO and many other national institutes - IIA; IUCAA; LEOS of ISRO; PRL; URSC of ISRO; VSSC of ISRO. We acknowledge the use of data from the Aditya-L1, the first solar mission of the ISRO, archived at the ISSDC.The AIA data used here are courtesy of SDO (NASA) and the AIA and HMI consortium. The GOES 8-16 X-ray/ magnetic field/ particle data are produced in real-time by the NOAA Space Weather Prediction Center (SWPC) and are distributed by the NOAA NGDC. Solar Orbiter is a space mission of international collaboration between ESA and NASA, operated by ESA. The STIX instrument is an international collaboration between Switzerland, Poland, France, Czech Republic, Germany, Austria, Ireland, and Italy. The STIX data used in this research is and necessary analysis tools are publicly available from STIX data center \citep{stix_d_cen}. We thank the open data policy of RSTN instruments. This research used version 6.0.5 \citep{sunpy_ver} of the SunPy open-source software package \citep{sunpy20} and PYTHON packages NumPy \citep{numpy}, Matplotlib \citep{matpltolib}, SciPy \citep{scipy}. SKS acknowledges funding from the European Research Council (ERC) under the European Union's Horizon 2020 research and innovation programme (grant agreement No. 101097844 — project WINSUN).

\end{acknowledgements}
\facilities{SDO(AIA), Solar Orbiter(STIX), GOES, Aditya-L1(SUIT), SOHO(Lasco), STEREO(WAVES), GONG, }
\appendix
\renewcommand\thefigure{\thesection\arabic{figure}}
\setcounter{figure}{0}
\section{Position of various spacecrafts}

\begin{figure}[ht!]
    \centering
    \includegraphics[trim={0cm 1cm 0cm 1cm}, clip, width=0.85\linewidth]{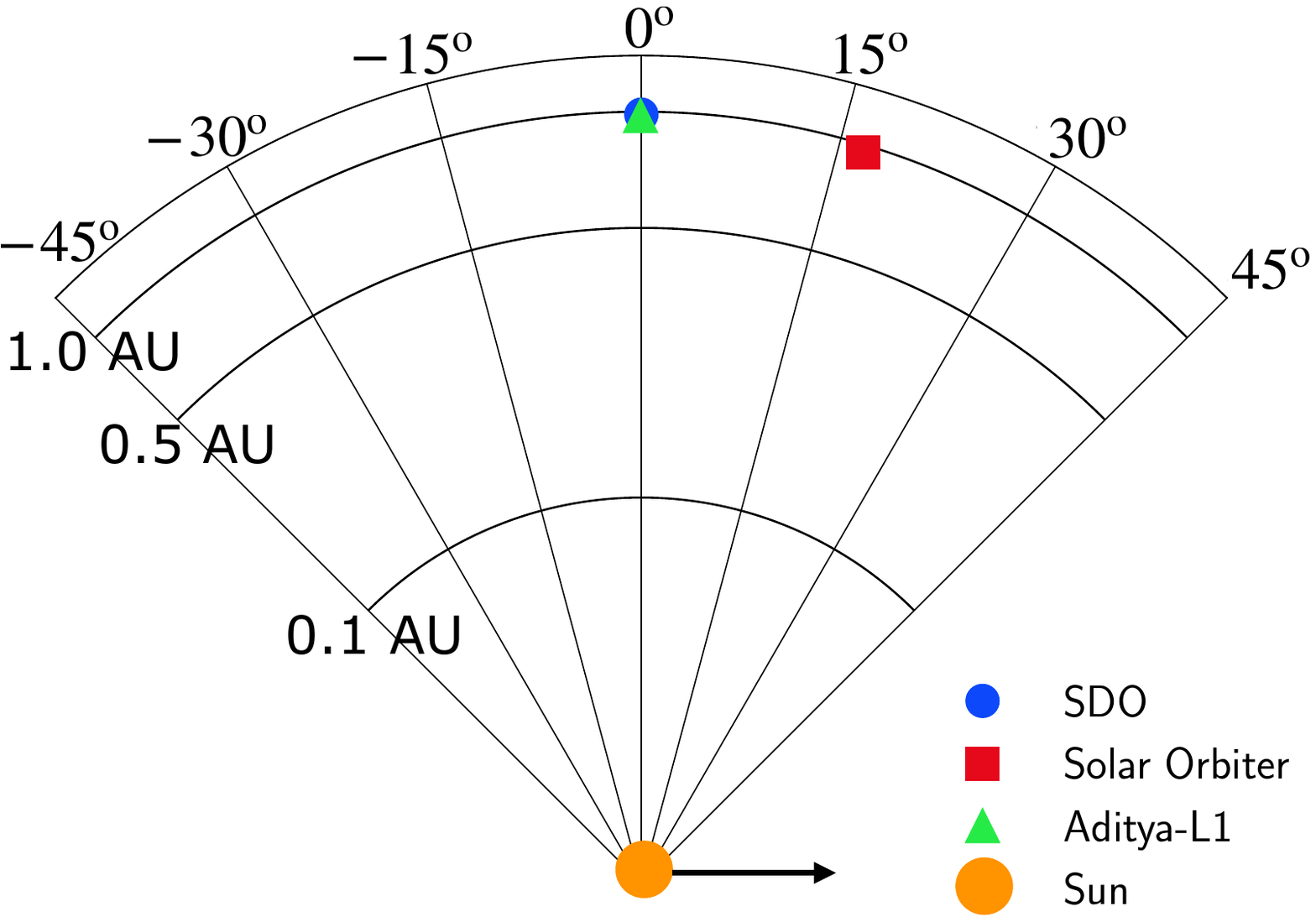}
    \caption{The position of various spacecraft with respect to the Sun at 21:39 UT on Dec 31, 2023 as labeled accordingly. Distance from the center of the Sun is in units of AU on a logarithmic scale. The solid black arrow shows the projected direction of the flare eruption and plasma blob propagation from AIA perspective.}
    \label{fig:sc_pos}
\end{figure}

In this study we have used observation from various instruments, which observe the Sun from two separate vantages, e.g. along the Sun Earth line and from Solar Orbiter vantage which was $\sim~\mathrm{16.8}$~{\degree} about the Sun Earth line. We show the positions of the spacecrafts in Fig.~\ref{fig:sc_pos} as marked.

\bibliography{mybib}{}
\bibliographystyle{aasjournal}
\end{document}